\documentclass[final,twoside]{IEEEtran}
\IEEEoverridecommandlockouts
\usepackage{cite}
\usepackage{url}
\usepackage{comment}
\usepackage{amsmath,amssymb,amsfonts}
\usepackage[linesnumbered,ruled,lined]{algorithm2e}
\usepackage{algpseudocode}
\usepackage{booktabs,caption}
\usepackage[flushleft]{threeparttable}
\usepackage{multirow}
\usepackage{mathtools}
\usepackage[pdftex]{graphicx}
\usepackage{textcomp}
\usepackage{mathtools}
\usepackage{dsfont}
\usepackage{subfig}
\usepackage{bm}
\usepackage{diagbox}
\usepackage{dblfloatfix} 
\usepackage[dvipsnames]{xcolor}

\makeatletter
\newcommand{\removelatexerror}{\let\@latex@error\@gobble}
\makeatother 
\DeclareMathOperator*{\argmax}{arg\,max}
\DeclarePairedDelimiter{\floor}{\lfloor}{\rfloor}
\DeclareMathOperator\arctanh{arctanh}
\def\BibTeX{{\rm B\kern-.05em{\sc i\kern-.025em b}\kern-.08em
    T\kern-.1667em\lower.7ex\hbox{E}\kern-.125emX}}
 \usepackage{setspace}
\makeatletter
\newcommand{\oset}[3][0ex]{%
  \mathrel{\mathop{#3}\limits^{
    \vbox to#1{\kern-2\ex@
    \hbox{$\scriptstyle#2$}\vss}}}}
\makeatother

\newcounter{relctr} 
\everydisplay\expandafter{\the\everydisplay\setcounter{relctr}{0}} 

\newcommand\labelrel[2]{%
  \begingroup
    \refstepcounter{relctr}%
    \stackrel{\textnormal{(\alph{relctr})}}{\mathstrut{#1}}%
    \originallabel{#2}%
  \endgroup
}

\newcommand{\algalign}[2]
{\makebox[\maxwidth][r]{$#1{}$}${}#2$}

\AtBeginDocument{\let\originallabel\label} 
\newcommand{\C}{\mathbb{C}}

\begin{document}

\title{Active Deep Decoding of Linear Codes}

\author{Ishay~Be'ery,
        Nir~Raviv,
        Tomer~Raviv,
        and~Yair~Be'ery,~\IEEEmembership{Senior~Member,~IEEE}
\thanks{This work was presented in part in the Future of Wireless Technology Workshop, Stockholm, Sweden, June 2019.}
\thanks{I. Be'ery, N. Raviv, T. Raviv and Y. Be'ery are with the School of Electrical Engineering, Tel-Aviv University, Tel-Aviv 6997801, Israel (e-mails: ishaybe@gmail.com, nirraviv89@gmail.com, tomerraviv95@gmail.com, ybeery@eng.tau.ac.il).}}

\markboth{Accepted to IEEE Transactions on Communications}
{Active Deep Decoding of Linear Codes}

\maketitle

\begin{abstract}
High quality data is essential in deep learning to train a robust model. While in other fields data is sparse and costly to collect, in error decoding it is free to query and label thus allowing potential data exploitation. Utilizing this fact and inspired by active learning, two novel methods are introduced to improve Weighted Belief Propagation (WBP) decoding. These methods incorporate machine-learning concepts with error decoding measures. For BCH(63,36), (63,45) and (127,64) codes, with cycle-reduced parity-check matrices, improvement of up to 0.4dB at the waterfall region, and of up to 1.5dB at the error-floor region in FER, over the original WBP, is demonstrated by smartly sampling the data, without increasing inference (decoding) complexity. The proposed methods constitutes an example guidelines for model enhancement by incorporation of domain knowledge from error-correcting field into a deep learning model. These guidelines can be adapted to any other deep learning based communication block. 
\end{abstract}


\IEEEpeerreviewmaketitle

\section{Introduction}
Decoding of error-correcting codes has changed over the last few years. The rise of machine-learning methods, primarily of the deep learning subset, changed the field significantly and comprehensively.

Nachmani et al. \cite{nachmani2016learning,nachmani2018deep}, proposed a model-based approach, placing learnable weights on Tanner graph's edges of the Belief Propagation (BP) algorithm \cite{pearl2014probabilistic} for linear codes. This approach is acknowledged as the original WBP, since had been the first in the field to learn a parameterized BP algorithm employing Stochastic Gradient Descent (SGD). The intuition offered was that the weights compensated for the short cycles in the Tanner graph. This addition improved the decoding performance. Lian et al. \cite{lian2018can} validated these results and further explored spatial and temporal weights sharing. Xu et al. \cite{xu2017improved, xu2018polar} generalized the method for both Tanner and factor graphs of polar codes. Considering model-free approaches, Gruber et al. \cite{gruber2017deep} proposed a fully connected (FC) neural networks (NN) approach, composed of linear and ReLU \cite{nair2010rectified} layers. This model achieved Maximum a-posteriori (MAP) performance on very short polar codes. Bennatan et al. \cite{bennatan2018deep} presented a combination of model-based and model-free approaches in which a NN was trained by the syndrome of the received message. Utilizing concurrent NN designs in addition with learning the code properties, via the composed syndrome, achieved performance improvement. Further contributions to the field lie in \cite{kim2018communication} and \cite{jiang2019deepturbo}, where neural decoders for convolutional codes were proposed together with befitting training methodologies. However, these methodologies impose substantial increase of complexity, at both training and decoding (inference). Specifically, \cite{kim2018communication} explores a recurrent neural networks (RNN) architecture for decoding while \cite{jiang2019deepturbo} focuses on an unconstrained novel structure requiring no knowledge of the BCJR (Bahl-Cocke-Jelinek-Raviv) algorithm.

As evident from the above recap, many researches paid great attention to the decoder's architecture, revealing a typical trend in the field. Yet another aspect of the mentioned decoding problem is the training data.  In \cite{nachmani2016learning,nachmani2018deep}, training over varying SNR (signal to noise ratio) ranges was explored. This leaded to different decoding performances over the same validation set. Regarding choice of a single optimal training point, Kim et al. \cite{kim2018communication} provides guidelines for choosing the best training SNR value. Gruber et al. \cite{gruber2017deep} showed that the choice of a training SNR value for generalization purposes is essential. A grid search is applied to locate the optimal single training SNR. This empirical result was followed by an analytical study in \cite{benammar2018optimal}. In the study, an entropy based analysis was performed, deriving a bound on the increase of the maximal error probability due to mismatched training and validation sets. The main conclusion is that no optimal training SNR for all validation sets \textit{exists}, but rather \textit{depends} on the specific validation data. One realization of this result is presented in \cite{lian2019learned}, where the WBP parameters are assumed to be SNR dependent. Multiple NN are used to infer the value for each parameter in the WBP algorithm at the validation phase, conditioned on the SNR.

Data is a vital part of deep learning methods, yet we see that it is not fully comprehended. Many researchers focus on preliminary choice of training data, followed by passive generation of examples during training. We rather search for an adaptive scheme which actively samples the training data to feed the neural decoder. Regarding complexity, \cite{lian2019learned} emphasizes that distribution-specific data requires unique analysis, but the additional NN cause extra complexity. In this paper we narrowed the view for schemes with no additional modules.

\noindent Our main contributions are:
\begin{enumerate}
    \item Active learning inspired approach is first applied, to our best knowledge, in the error-correcting codes field.
    \item Performance improvement with no decoding complexity penalty. 
    \item  Directing the effort of the machine-learning decoding community to data-tailored solutions.
\end{enumerate}

\noindent We call our approach \textit{active deep decoding}.

The paper is organized as follows. Section \ref{Preliminaries} covers notation and definitions. Section \ref{data_exploration} explores different decoding parameters and sets the ground for the novel methods. Section \ref{active_learning} introduces a detailed explanation of the methods. Section \ref{results} presents experiments and results and section \ref{conclusions} concludes the paper.

\section{Preliminaries} \label{Preliminaries}
\subsection{Notation}

\begin{figure*}[t!]
\centering
\includegraphics[width=\textwidth]{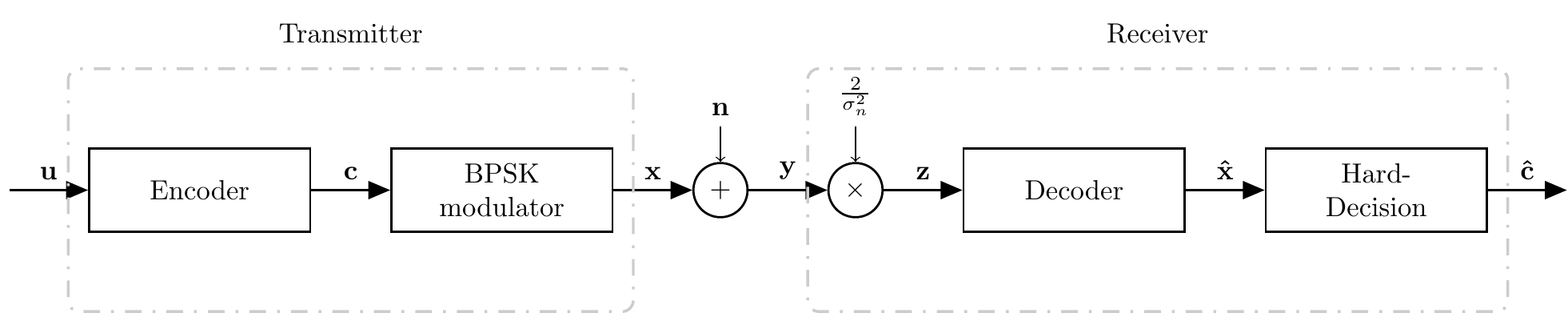}
\caption{System Diagram}
\label{fig:system_diagram}
\end{figure*}

We denote scalars in \textit{italics} letters and vectors in $\mathbf{bold}$. Capital and lowercase letters stand for a random vector and it's realization, respectively. For example, $\mathbf{C}$ and $\mathbf{c}$ stand for the codeword random vector and it's realization vector. $\mathbf{X}$ and $\mathbf{Y}$ are the transmitted and received channel words. $\mathbf{\hat{X}}$ denotes the decoded modulated-word, while $\mathbf{\hat{C}}$ denotes the decoded codeword. The $i^{th}$ element of a vector $\mathbf{v}$ will be denoted with a subscript $v_i$. 

We will only deal with the AWGN channel in this work, denoting the SNR by $\rho$ for convenience. We also denote the code by $\C$, with minimum Hamming distance $d_{min}$ and code length $N$. Let $\mathbf{u}$, $\mathbf{x}$ and $\mathbf{y}$ denote the message word, the transmitted word (after encoding and BPSK modulation) and the received word (with Gaussian noise $\mathbf{n} \sim \mathcal{N}(\mathbf{0},\,\sigma_n^{2}\mathbf{I})$) respectively, see Figure \ref{fig:system_diagram}. Note that one always decodes the received LLR word $\mathbf{z}$, not $\mathbf{y}$. Let $dist(\mathbf{c}_1,\mathbf{c}_2)$ denote the Hamming distance between two codewords $\mathbf{c}_1$ and $\mathbf{c}_2$. Specifically, we denote by $d_H$ the Hamming distance between the encoded codeword $\mathbf{c}$ and the decoded word $\hat{\mathbf{c}}$. The received word will always be decoded correctly by a hard-decision decoder if the Hamming distance between $\mathbf{c}$ and $\mathbf{y}$ demodulated by hard-decision is $t_H=\floor{\frac{d_{min}-1}{2}}$ at most. Let $T$ be a latent binary variable \cite{goodfellow2016deep}, which denotes successful decoding of the $NN$ decoder, with a value of 1 if $\mathbf{c}=\mathbf{\hat{c}}$ (similarly $d_H=0$).

At last we denote $I(X;Y)$ as the mutual information between the two random variables, $X$ and $Y$.

\subsection{Training by Different Parameters}

Let $\Gamma_{\theta}(S)$ be a distribution over received words $\mathbf{Y}$, parameterized by hyperparameters $\theta \in \Theta$ set with values $S$. For example, let $\theta$ be $\rho$ and $S = 1$dB. Then, a training sample is drawn (assuming the all-zero codeword is transmitted) according to $P_\mathbf{Y} (\mathbf{y};\rho = 1)$. For a batch of i.i.d. training samples, the entire sampling procedure is repeated $n$ times, where $n$ is the required batch size and both $\theta$ and $S$ may vary in the same batch. We denote a batch sampled according to $\Gamma$ by $\mathbf{y_{\gamma}}$.

\subsection{Weighted BP Decoding}
 The Belief Propagation (BP) is an inference algorithm used to calculate the marginal probabilities of nodes in a graph efficiently. Pearl \cite{pearl2014probabilistic} also advocated the utilization of this algorithm for graphs with loops, along with a remark that it is an approximation only. This version is called the loopy belief propagation. A full derivation from the general case to linear codes can be found in \cite{richardson2008modern}. We provide main details next. 

The Tanner graph is an undirected graphical model, constructed of nodes and edges. The nodes are of two types - variables and checks nodes. A variable node corresponds to a single bit of the received codeword. Each check node corresponds to a row in the code's parity check matrix. An edge exists between a variable $v$ and a check node $h$ iff variable $v$ participates (has coefficient 1) in the condition defined by the $h^{th}$ row in the parity check matrix.

\noindent The initialization of the variable nodes:
\begin{equation*}
    z_v = \log\frac{P(c_v=0|y_v)}{P(c_v=1|y_v)} = \frac{2y_v}{\sigma_n^2}
\end{equation*}
The subscript $v$ indicates a variable node and $z$ stands for a LLR (log-likelihood ratio) value. The last equality is true for AWGN channels with common BPSK mapping to $\{\pm1\}$.

The message passing algorithm proceeds by iteratively passing messages over edges from variables nodes to check nodes and vice versa. The BP message from node $a$ to node $b$ at iteration $i$ will be denoted by $m_{i,(a,b)}$ with the convention that $m_{0,(a,b)} = 0$ for all $a$,$b$ combinations. Variable-to-check messages are updated in odd iterations according to the rule:
\begin{equation*}
    m_{i,(v,h)}= z_v + \sum_{\substack{(h',v), h' \neq h}} m_{i-1,(h',v)}
\end{equation*}
While the check-to-variable messages are updated in even iterations by:
\begin{equation*}
    m_{i,(h,v)}= 2\arctanh{ \left(\prod_{(v',h),v' \neq v} \tanh \left(\frac{m_{i-1,(v',h)}}{2}\right)\right)}
\end{equation*}
Finally, the output variable node value is calculated by:
\begin{equation*}
    \hat{x}_v = z_v + \sum_{\substack{(h',v), h' \neq h}} m_{2\tau,(h',v)}
\end{equation*}
Where $\tau$ is the number of BP iterations and all values considered are LLR values. In \cite{nachmani2016learning,nachmani2018deep}, learnable weights are assigned to the variable-check message passing rule: 
\begin{equation} \label{eq:v2c_layer}
    m_{i,(v,h)}= \tanh \left(\frac{1}{2}\left(w_{i,v}z_v+\sum_{\substack{(h',v) \\ h' \neq h}} w_{i,(h',v,h)}m_{i-1,(h',v)}\right)\right)
\end{equation}
And to the output marginalization:
\begin{equation} \label{eq:output_layer}
    \hat{x}_v\!=\!\sigma\left(\!-\!\left[w_{2\tau+1,v}z_v \!+\! \sum_{\substack{(h',v) \\ h' \neq h}} w_{2\tau+1,(h',v)}m_{2\tau,(h',v)}\right]\right)
\end{equation}
where $\sigma$ is the sigmoid function. We denote by $\mathbf{w} = \{w_{i,v},w_{i,(h',v,h)},w_{i,(v,h')}\}$ the set of weights. 
Note that no weights are assigned to the check-variable rule, which now takes the form:
\begin{equation} \label{eq:c2v_layer}
    m_{i,(h,v)}= 2\arctanh{ \left(\prod_{(v',h),v' \neq v} m_{i-1,(v',h)}\right)}
\end{equation}

This decision is explained by expected numerical instabilities due to the arctanh domain.

This formulation unfolds the loopy algorithm into a NN. One can see that the hyperbolic tangent function was moved from check-variable rule to scale the message to a reasonable output range. A sigmoid function is used to scale the LLR values into the range $[0,1]$. An output value in the range $(0.5,1]$ is considered a '1' bit, otherwise a '0' bit (value of $0.5$ was attributed to the '0' bit randomly). Training is done with the Binary Cross Entropy (BCE) multiloss: 

\begin{equation*}
\begin{split}
    L(\mathbf{\mathbf{c},\hat{x}}) = -\frac{1}{N}\sum_{t=1}^{\tau}\sum_{v=1}^{N} [c_v\log{\hat{x}_{v,t}} + (1-c_v) \log{(1 - \hat{x}_{v,t})}]
\end{split}
\end{equation*}

For a comprehensive explanation of the subject, please refer to \cite{nachmani2016learning,nachmani2018deep}.

\section{Data Exploration}\label{data_exploration}

We start exploring the data with a question in mind - \textit{do all words contribute equally to the neural training?}

\subsection{The SNR Parameter - A Motivation}

We inspect how possessing the knowledge of $\rho$ can affect the training data and model choices.

Regarding training data, Gruber et al. \cite{gruber2017deep} trains multiple neural decoders, each decoder trained with data drawn from $\Gamma_{\rho}(i)$ where $-4 \leq i \leq 8, i\in\mathbb{Z}$. The $NVE(\rho_t,\rho_v)$ measure is suggested in \cite{gruber2017deep} to compare between the trained models. One can notice that the model diverges when trained over only correct or noisy words, drawn from high or low SNR, respectively. In \cite{kim2018communication} guidelines for choosing $\rho_t$ are provided. The value is chosen so that the neural decoder's training set is comprised from $\mathbf{y}$ near the decision boundary.

Regarding model choices, a hidden assumption of \cite{lian2019learned} is that $\mathbf{y_{\gamma}}$ which are drawn from $\Gamma_{\rho}(S_1)$ and $\Gamma_{\rho}(S_2)$ ($S_1\neq S_2$) require different decoder weights, $\mathbf{w_1},\mathbf{w_2}$. One may observe that knowledge possession of $\rho_v$ is also mandatory for all LLR-based decoders (as an estimate is required to compute LLRs). It is quite straightforward to show that the next mutual information inequality holds: 

\begin{equation*}
\label{mutual_info}
I(\mathbf{Y},\rho_v;T) \labelrel={mutual_info:a} I(\mathbf{Y};T) + I(\rho_v;T|\mathbf{Y}) \labelrel\geq{mutual_info:b} I(\mathbf{Y};T)
\end{equation*}

where ~\eqref{mutual_info:a} follows from the mutual information chain rule, and ~\eqref{mutual_info:b} follows from the non-negativity of mutual information. Thus, the additional information of $\rho_v$ can only aid decoding. This information of the channel and the decoder distributions, conditioned on the received word, may be non-zero for sub-optimal decoders. In \cite{lian2019learned}, inference does not only require $\rho_v$ knowledge, but is also $\rho_v$ dependent. In other words - the model is data dependent.

\subsection{Objective Formulation}\label{sub:formulation}

Motivated by the above discussion, our main goal is to find parameters other than the SNR, which define a new $\Gamma$, $\Gamma_{new}$. We want that training the WBP over $\Gamma_{new}$ will achieve as high decoding performance as possible.

Let $\kappa$ denote the contribution of a word, in the training phase, to the validation decoding performance. We associate higher contribution words with higher $\kappa$ value. Our goal is to find parameters $\theta\in\Theta$ and corresponding values $S$ defining words distribution $\Gamma_{\theta}(S)$ such that the $\kappa$ value integrated over the distribution is maximized:

\begin{equation*}
    \argmax_{\theta,S}\int\displaylimits_{\mathclap{\mathbf{y}\in\Gamma_{\theta}(S)}} \kappa(\mathbf{y}) 
\end{equation*} 

The solution to this equation is intractable due to the infinite number of such parameters and values, thus we seek heuristic-based solution. We choose the parameters based on the vast decoding knowledge while using the above insights. In particular, $\mathbf{y_{\gamma}}$ should be neither too noisy nor absolutely correct and should lie close to the decision boundary. Recall that throughout the paper we use the AWGN channel. Therefore, we search for parameters $\theta'$ which limit the feasible $\mathbf{y_{\gamma}}$ of the channel distribution $\Gamma_{\rho}(S)$, associated with $K_{\rho}(S)$, to $\Gamma_{\rho,\theta'}(S,A)$, associated with higher $K_{\rho,\theta'}(S,A)$, where we denote $K_{\theta}(S) = \int\limits_{\mathbf{y}\in\Gamma_{\theta}(S)}\kappa(\mathbf{y})$.

\subsection{Distance Parameter}\label{sub:distance}

Some received words are undecodable due to the locality of the decoding algorithm, the Tanner graph structure induced by the parity-check matrix or a high Hamming distance. By sampling from specific  $\Gamma_{\rho,d_H}(S,A)$ one can easily control the number of erroneous bits in $\mathbf{y}$. Choosing such words with a reasonable Hamming distance between them and the transmitted words decreases the amount of undecodable words in $\Gamma$. 

To justify the above claims we trained a WBP decoder without any correct received words, $d_H$=0, and without high noise words, $d_H>t_H$. The training setup is similar to the one used in section \ref{results}. The results show an improvement of up to 0.5dB by sampling according to this simple scheme, confirming our intuitions. By drawing data according to a distribution, and not according to the SNR, we have further control on training words’ properties. We elaborate more on this subject in \ref{sub:distanceApproach}.

\textit{With this short experiment we manage to answer the question we set to ask - do all words contribute equally to the training?} A definitive answer is \textbf{no}.
\raggedbottom 
\subsection{Reliability Parameters}\label{sub:reliability}

Soft in soft out (SISO) decoding compose the received signal to n LLR values, $\{z_1,\ldots,z_n\}$. In general $z_v \in (-\infty,\infty)$ but in practice we limit their value by choosing appropriate threshold. The closer the $z_v$ to 0, the less reliable it is. We consider mapping the LLR values to bits in two steps. First mapping LLR values to probabilities:
\begin{equation*}
    \Pi_{LLR \rightarrow Pr}(z_i)= \sigma(-z_i)
\end{equation*}
The next rule maps probability into corresponding bit: 
\begin{equation*}
    \Pi_{Pr \rightarrow bit}(\widetilde{z}_i)= 
    \begin{cases}
          1, & \text{if}\ \widetilde{z}_i>0.5 \\
          0, & \text{otherwise}
    \end{cases}
\end{equation*}
The process of direct quantization from LLR to bits is called hard decision (HD) decoding:
\begin{equation}
    \Pi_{HD}(z_i)=\Pi_{Pr \rightarrow bit}(\Pi_{LLR \rightarrow Pr}(z_i))
\end{equation}
Obviously there is information loss in the process:
\begin{equation*}
    \Pi_{HD}(z_1) = \Pi_{HD}(z_2) \\ \nRightarrow z_1=z_2
\end{equation*}

We seek numeric parameters which quantify reliability of a given $\mathbf{z}$. Two parameters that we inspected and found fitting to the task are defined below:

\textbf{Average Bit Probability} - the deviation of the channel output probabilities from the corresponding transmitted bits:
\begin{equation} \label{eq:ABP}
    \eta_{ABP} (c_i,z_i)= \frac{1}{N}\sum_{i=1}^{N} |c_i-\Pi_{LLR \rightarrow Pr}(z_i)|
\end{equation}

\textbf{Mean Bit Cross Entropy} - this parameter quantifies how close are the two probability distributions at the transmitter and at the receiver (before decoding):
\begin{equation} \label{eq:MBCP}
\begin{split}
    \ell_{MBCE} (c_i,z_i)= \frac{1}{N}\sum_{i=1}^{N} |c_i\cdot\log(\Pi_{LLR \rightarrow Pr}(z_i)) +\\+(1-c_i)\cdot\log(1-\Pi_{LLR \rightarrow Pr}(z_i))|.
\end{split}
\end{equation}

By limiting the distribution to $\Gamma_{\rho, \eta_{ABP}, \ell_{MBCE}}(S,A_1,A_2)$, we have a better control of the distribution of $\mathbf{y}$, and consequently of $\mathbf{z}$, such that $\mathbf{y_{\gamma}}$ has higher $\kappa$ on average. The intuition guiding us, again, is that higher $\kappa$ words lie close to the decision boundaries. Referring to \ref{sub:formulation}, we need to choose $A_1,A_2$ such that $K_{\rho, \eta_{ABP},\ell_{MBCE}}(S,A_1,A_2)$ is maximized.

\subsection{Correlation with SNR}\label{sub:correlation}

\begin{figure}[t]
\centering
\includegraphics[width=8cm, height=7cm]{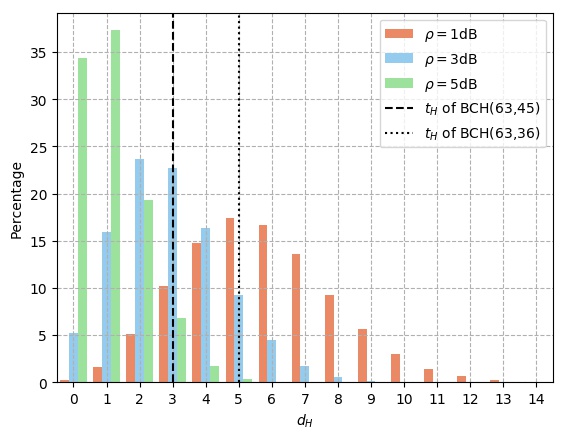}
\caption{Distribution of $d_H$ by $\rho$}
\label{fig:d_vs_snr}
\end{figure}

\begin{figure}[t]
\centering
\includegraphics[width=8cm, height=7cm]{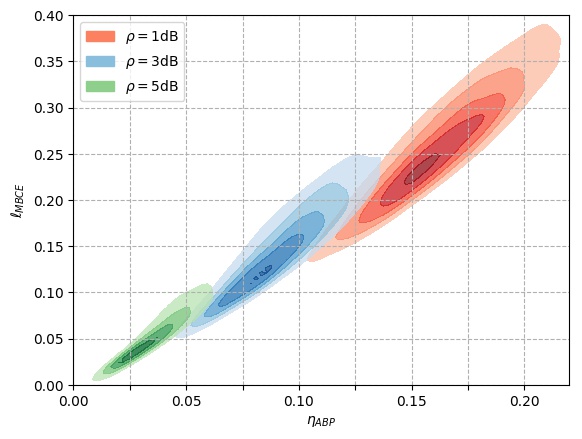}
\caption{Distribution of reliability by $\rho$}
\label{fig:rel_vs_snr}
\end{figure}

Figures \ref{fig:d_vs_snr} and \ref{fig:rel_vs_snr} show the correlation of the above parameters to $\rho$ and $T$. In both figures 100,000 codewords were simulated per $\rho$ on code with length of 63 bits. Regarding Figure \ref{fig:d_vs_snr}, one can see that each $\rho$ defines a different probability distribution of $d_H$ values. This figure is unique for each code length and simulated $\rho$. The higher the SNR - the lower the $d_H$ center of this probability distribution. High $\rho$ includes high amount of no errors frames, while low $\rho$ value induces lots of high noise received words with $d_H$ higher than $t_{H}$. Both $t_H$ values for the two codes BCH(63,36) and BCH(63,45) are also plotted on this figure. Figure \ref{fig:rel_vs_snr} represents similar notion in regard to reliability. Each $\rho$ defines a probability distribution over the two parameters so that the higher the $\rho$ is, the closer the distribution is to the origin. Here we do not have a defined threshold for correct and highly incorrect words, $\mathbf{y}$, as before, thus we must sample from this probability distribution much more carefully.

One thing we ignored so far is the evolution of the decoder during training. Obviously, as the decoder trains its' decision regions are altered - changing the optimal $\theta,S$ to sample by. In order to train the decoder with $\mathbf{y}$ close to the decision boundaries at every stage, the distribution $\Gamma_{\theta}(S)$ we draw from must change actively during the training. A known method in machine-learning field for doing so is called - \textit{active learning}.

\section{Active Learning} \label{active_learning}
Active learning is a supervised learning method, which deals with an oracle that actively chooses the samples from a large pool of unlabelled data to feed the model. The oracle can be human annotator or a machine based one. Two important questions regarding this process are "why is active learning used" and "how is a batch queried". The solution to the former question is straightforward - the reason for using this method is the queried batch is assumed to benefit the training of the model more than using a random training batch, on average. The solution to the second question is achieved by introducing a metric which shows informativeness. At each training step, the batch with the highest metric value is considered the most informative thus it is queried. It is widely used in medical systems and in situations when annotating data is expensive, thus training data must be chosen punctiliously. For additional information on active learning see \cite{settles2009active,fu2013survey}.

In a stream based approach batches are generated one by one. A selective sampling approach is one in which the data to be queried is selected based on some metric.
An underlying assumption in active learning stream-based selective sampling approach is that data is free to obtain. In our error-correcting codes domain data is unlimited when the channel model is known (as AWGN) or can be fairly easily collected when channel model is unknown, and we do not need to annotate it by hand. This is a huge advantage and a strong claim in favor of using this method in decoding. 
Traditionally, this method is used on unlabeled stream or pool of data. In this case one would want to choose which samples are worth labeling and training. In our case all samples are labeled. Therefore, our goal is to perform the training procedure with the highest
$\kappa$ for the received words. 

We hereby present the main two active learning approaches taken.

\subsection{Stream-Based Selective Sampling by Hamming Distance}\label{sub:distanceApproach}

\begin{figure}[t!]
 \removelatexerror
  \begin{algorithm}[H]
    \caption{Stream-based sampling by Hamming distance}
    \label{alg:SampleByDistance}
    \SetAlgoLined
    \SetKwProg{SampleByDistance}{SampleByDistance}{}{}
    \SetKwInOut{Initialization}{Initialization}
    \Initialization{$DEC$ as in \cite{nachmani2016learning}}
    \SetKwInOut{Input}{Input}
    \SetKwInOut{Output}{Output}
    \Input{current decoder $DEC$ \newline
    $S=\{s_1,\ldots.,s_n\}$ set of SNR values \newline
    $A=\{1,\ldots.,d_{max}\}$ set of $d_H$ values \newline $\mathbf{c}$ encoded word}
    \Output{improved model $DEC$}
    \SampleByDistance{$(DEC,S,A,\mathbf{c})$}{
        \While{error decreases}{%
             sample batch $Q$ from $\Gamma_{\rho,d_H}(S,A)$\;
            \For{$\mathbf{y}$ in $Q$}{%
                $d_{in} \gets dist(\Pi_{HD}(\mathbf{y}),\mathbf{c})$\;
                $d_{out} \gets dist(\mathbf{\hat{c}},\mathbf{c})$\;\label{line:forward}
                \uIf{$d_{out}=0$ or $d_{out} \geq d_{in}$}{ 
                    $Q \gets Q \setminus \mathbf{y}$; \label{line:filter}}
                }
            $DEC \gets$ update model based on $Q$;\label{line:update}
        }
        \KwRet{$DEC$}\;
    }
    \end{algorithm}
\end{figure}

The first approach is presented in Algorithm \ref{alg:SampleByDistance}, where at each time step, the current neural model (line \ref{line:forward}) determines the next queried batch (line \ref{line:filter}) for the model update (line \ref{line:update}). This algorithm is based on intuitions from Subsection \ref{sub:distance}, remove successfully decoded $\mathbf{y}$ in addition to very noisy $\mathbf{y}$ from training (lines 7-\ref{line:filter}). These received words are far from the decision boundary thus harm training. Why these $\mathbf{y}$ can harm the training can also be explained from the learning signal perspective. On one hand, the real signal is nearly impossible to be recovered from a very noisy $\mathbf{y}$, thus the learning signal towards a minima is very low. On the other hand, for very reliable $\mathbf{y}$, the learning signal is low, since for every direction of decision the model takes these reliable words will be decoded successfully. Thus, they are not informative for learning. 

\begin{figure}[t!]
 \removelatexerror
  \begin{algorithm}[H]
    \caption{Stream-based sampling by reliability parameters}
    \label{alg:SampleByReliability}
    \SetAlgoLined
    \SetKwInOut{Initialization}{Initialization}
    \Initialization{$DEC$ as in \cite{nachmani2016learning}}
    \SetKwInOut{Input}{Input}
    \SetKwInOut{Output}{Output}
    \SetKwProg{SampleByReliability}{SampleByReliability}{}{}
    \Input{current decoder $DEC$ \newline $S=\{s_1,\ldots.,s_n\}$ set of SNR values \newline $b$ desired batch size \newline $\mathbf{c}$ encoded word} 
    \Output{improved model $DEC$}
    \SampleByReliability{$(DEC,S,m,\mathbf{c})$}{
        $\boldsymbol{\mu},\boldsymbol{\Sigma} \gets$ ChoosePrior($S$,$\mathbf{c}$)\;  \label{line:prior}
        \While{error decreases}{%
            sample batch $Q$ from $\Gamma_{\rho}(S)$\;
            $\eta_{ABP} \gets$ calculate by equation $(\ref{eq:ABP})$ per word\;
            $\ell_{MBCE} \gets$ calculate by equation $(\ref{eq:MBCP})$ per word\;
            $\boldsymbol{\theta} \gets [\eta_{ABP}, \ell_{MBCE}]$\;
            $\boldsymbol{w} \gets f(\boldsymbol{\theta}|\boldsymbol{\mu}, \boldsymbol{\Sigma})$\;
            $\boldsymbol{\tilde{w}} \gets \boldsymbol{w}/||\boldsymbol{w}||_1 $\;  \label{line:normalize}
            $\tilde{Q} \gets$ random sampling $b$ words from $Q$ w.p ${\boldsymbol{\tilde{w}}}$\; \label{line:sampling}
            $DEC \gets$ update the model based on $\tilde{Q}$;
        }
        \KwRet{$DEC$}\;
   }
   \end{algorithm}
\end{figure}

\begin{figure}[t!]
 \removelatexerror
  \begin{algorithm}[H]
    \caption{Choose prior}
    \label{alg:ChoosePrior}
    \SetAlgoLined
    \SetKwInOut{Input}{Input}
    \SetKwInOut{Data}{Data}
    \SetKwInOut{Output}{Output}
    \SetKwProg{ChoosePrior}{ChoosePrior}{}{}
    \Input{$S=\{s_1,\ldots.,s_n\}$ set of SNR values\newline
    $\mathbf{c}$ encoded word} 
    \Data{$\tau_{set}=\{\tau_1,\ldots,\tau_r\}$ set of iterations \newline
           $\epsilon_{set}=\{\epsilon_1,\ldots,\epsilon_{r+1}\}$ set of colors}
    \Output{prior distribution parameters \boldsymbol{$\mu$},\boldsymbol{$\Sigma$}}
    \ChoosePrior{$(S,\mathbf{c})$}{
        sample batch $Q$ from $\Gamma_{\rho}(S)$\;
        $\eta_{ABP} \gets$ calculate by eq. (\ref{eq:ABP})\;
        $\ell_{MBCE} \gets$ calculate by eq. (\ref{eq:MBCP})\;
        \For{$\tau_i$ in $\tau_{set}$}{%
            $DEC \gets$ construct BP with $\tau_i$ iterations\;
            \For{$\mathbf{y}$ in $Q$}{%
                $d \gets dist(\mathbf{\hat{c}},\mathbf{c})$\;
                \uIf{$d=0$}{%
                    Plot point ($\eta_{ABP}$,$\ell_{MBCE}$) in color $\epsilon_i$\;
                }
                \uElse{Plot point ($\eta_{ABP}$,$\ell_{MBCE}$) in color $\epsilon_{i+1}$\;}
            }
        }
        $\boldsymbol{\mu},\boldsymbol{\Sigma} \gets$ set empirically on decodable words\;
        \KwRet{$\boldsymbol{\mu},\boldsymbol{\Sigma}$}\;
    }
    \end{algorithm}
\end{figure}

\subsection{Stream-Based Selective Sampling by Reliability Parameters}
The second approach we present exploits the reliability of a given $\mathbf{y}$, see Algorithm \ref{alg:SampleByReliability}. Inspired by the common uncertainty sampling query framework, we first calculate $\Gamma_{\rho,\eta_{ABP},\ell_{MBCE}}(S,A_1,A_2)$ for several untrained BP decoders with different number of iterations $\tau_{set}=\{\tau_1,\ldots,\tau_r\}$ empirically. We chose to query each batch by setting a prior on $\eta_{ABP},\ell_{MBCE}$. We elaborate on the prior and batch selections. Firstly, the prior was chosen as a Normal distribution with expectation, $\bm{\mu}$, and covariance matrix, $\bm{\Sigma}$, over $\mathbf{y}$ that are decodable by adding iterations to the standard BP decoder. The prior selection is summarised in Algorithm \ref{alg:ChoosePrior}. These $\mathbf{y}$ are assumed to be close to the decision boundaries, since BP decoders with additional iterations are able to decode them. We want the WBP to compensate for these additional iterations by training. Secondly, in Algorithm \ref{alg:SampleByReliability}, the batch was queried by performing a few trivial steps (lines \ref{line:prior}-\ref{line:normalize}). The last step (line \ref{line:sampling}) includes random sampling of a given size batch by the normalized weights as the probabilities, without replacement.

One important note is that the uncertainty sampling method is usually performed over the neural model output signal, while here we use it over the input signal. That is because the multiple BP decoders are the baseline for improvement, not the weighted decoder. 

\section{Experiments and Results} \label{results}

We present the results of training and applying the approaches mentioned in \ref{active_learning} for three different linear codes BCH(63,45), BCH(63,36), BCH(127,64) with $t_H=3$, $t_H=5$ and $t_H=10$, respectively. We use the cycle-reduced (CR) parity-check matrices as appear in \cite{channelcodes}, thus evaluating our method when the number of short cycles is already small and improvement by altering weights is harder to achieve. The number of iterations is chosen as 5 as in \cite{nachmani2016learning,nachmani2018deep,lian2018can,xu2017improved,xu2018polar,bennatan2018deep}, who set a benchmark in the field. The zero codeword is used for training, due to symmetry, as in \cite{nachmani2016learning,nachmani2018deep}. It also serves as the codeword in Algorithms \ref{alg:SampleByDistance} and \ref{alg:SampleByReliability}. All other training relevant hyperparameters are summarised in Table I. All WBP decoders are trained until convergence. We apply two methods - Hamming distance based and reliability based. Regarding the active learning hyperparameters, for the distance approach, and in order to stay consistent, we chose the same  $d_{max}$ for the two short codes. All hyperparameters are summarised in Table II. As a follow-up to section \ref{sub:distance}, we also apply $d_H$ filtering to the reliability method. This is referred as the reliability \& $d_H$ filtering in Table II.

\renewcommand{\arraystretch}{1.1} 
\begin{table}[t]
    \centering
      \begin{threeparttable}
        \caption{Training Hyperparameters}
          \label{tab:train}
            \begin{tabular}{| c  c |}
                \hline
                \textbf{Hyperparameters} & \textbf{Values}\\ 
                \hline
                Architecture & Feed Forward\\ 
                \hline 
                Initialization & as in [1] (*) \\
                \hline
                Loss Function & BCE with Multiloss \\
                \hline
                Optimizer & RMSPROP \\
                \hline
                $\rho_t$ range & 4dB to 7dB \\  
                \hline
                Learning Rate & 0.01 \\
                \hline
                Batch Size & 1250 / 300 words per SNR (**) \\
                \hline
                Messages Range & $(-10,10)$ \\
                \hline
            \end{tabular}
            \begin{tablenotes}
              \small
              \item (*) $w_{i,v}$ in eqs. (1) and (2) set to constant 1 since no additional improvement was observed. 
              \item (**) for the 63 / 127 code length, respectively. 
            \end{tablenotes}
      \end{threeparttable}
\end{table}

\begin{table*}
    \centering
      \begin{threeparttable}
        \caption{Active Learning Hyperparameters}
        \label{tab:active}
            \begin{tabular}{ | c | c | c | c | }
                \hline
                \textbf{Method} & \textbf{Hyperparameters} & \textbf{CR-BCH N=63} & \textbf{CR-BCH N=127}\\ 
                \hline
                Hamming distance & $d_{max}$ & 2 & 4 \\ 
                \hline 
                \multirow{3}{*}{Reliability} & $\tau_{set}$ & \multicolumn{2}{c|}{$\{5,7,10,15\}$}\\
                \cline{2-4}
                 & $\boldsymbol{\mu}$ & $(0.025,0.1)$ &  $(0.03,0.1)$\\
                \cline{2-4}
                 & $\boldsymbol{\Sigma}$ & \multicolumn{2}{c|}{$\begin{bmatrix} 6.25\cdot10^{-4} & 0 \\ 0 & 5.625\cdot10^{-3}  \end{bmatrix}$}\\
                 \hline
                 \multirow{4}{*}{Reliability \& $d_H$ filtering} & $d_{max}$ & 3 & 5 \\
                 \cline{2-4}
                 & $\tau_{set}$ & \multicolumn{2}{c|}{$\{5,7,10,15\}$}\\
                 \cline{2-4}
                 & $\boldsymbol{\mu}$ & $(0.025,0.1)$ &  $(0.03,0.1)$\\
                 \cline{2-4}
                 & $\boldsymbol{\Sigma}$ & \multicolumn{2}{c|}{$\begin{bmatrix} 6.25\cdot10^{-4} & 0 \\ 0 & 5.625\cdot10^{-3}  \end{bmatrix}$}\\
                \hline
            \end{tabular}
      \end{threeparttable}
\end{table*}

\begin{figure*}[!t]
\centering
\subfloat[CR-BCH(63,36)]{
            \includegraphics[width=0.45\textwidth]{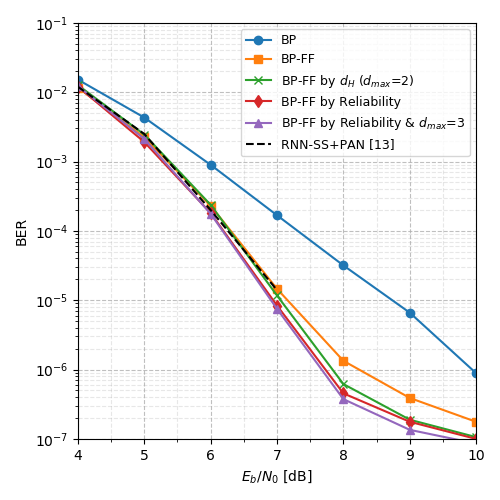}
            \includegraphics[width=0.45\textwidth]{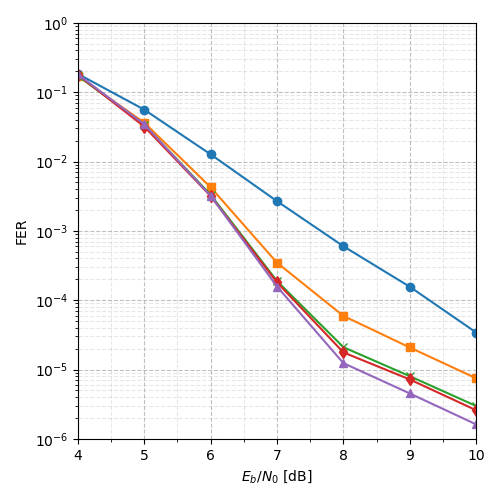}
            \label{subfig:63_36}
          }

\subfloat[CR-BCH(63,45)]{
            \includegraphics[width=0.45\textwidth]{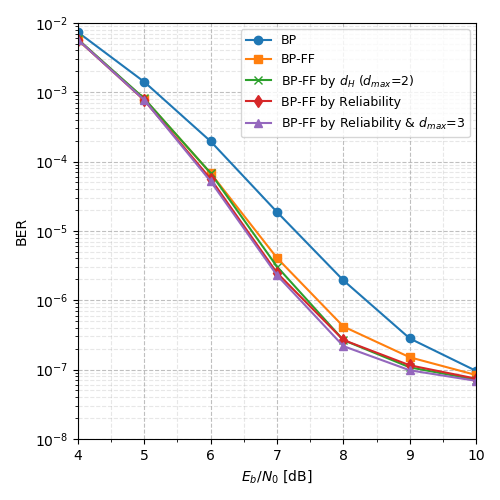}
            \includegraphics[width=0.45\textwidth]{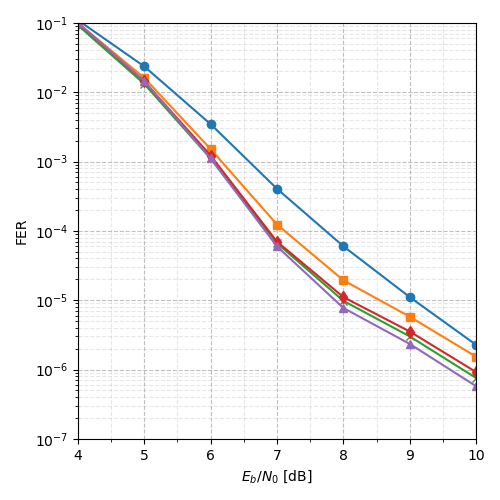}
            \label{subfig:63_45}
          }
\phantomcaption
\end{figure*}   

\begin{figure*}[!t]
    \centering
    \ContinuedFloat
    \subfloat[CR-BCH(127,64)]{
                \includegraphics[width=0.45\textwidth]{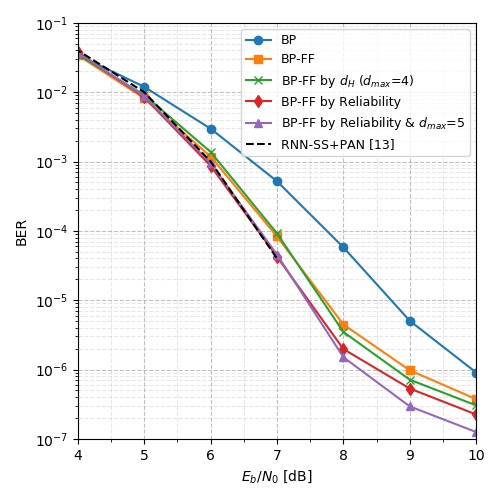}
                \includegraphics[width=0.45\textwidth]{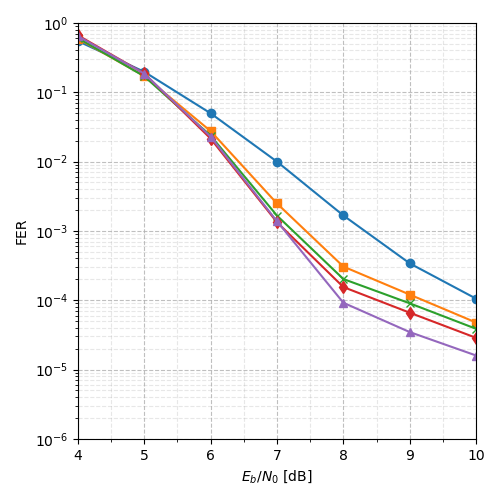}
                \label{subfig:127_64}
    }
    
    \caption{BER and FER comparison: ($\textcolor{NavyBlue}{\bullet}$) Regular BP, ($\textcolor{BurntOrange}{\blacksquare}$) BP-FF \cite{nachmani2016learning}, ($\textcolor{Green}{\times}$) Hamming distance approach, ($\textcolor{BrickRed}{\blacklozenge}$) Reliability approach, ($\textcolor{Fuchsia}{\blacktriangle}$) Reliability approach with filter by $d_H$}
    
    \label{fig:performance}
\end{figure*}

\begin{table*}[t]
    \centering
      \begin{threeparttable}
        \caption{Best Decoding  Gains (*)(**)}
        \label{tab:gain}
            \begin{tabular}{ | c | c | c | c | c | }
                \hline
                \backslashbox{\textbf{Code}}{\textbf{Region}} & \multicolumn{2}{|c|}{\textbf{Waterfall}} & \multicolumn{2}{|c|}{\textbf{Error-floor}}\\ 
                \hline
                & \textbf{BER[dB]} & \textbf{FER[dB]} & \textbf{BER[dB]} & \textbf{FER[dB]} \\ 
                \hline 
                \textbf{CR-BCH(63,36)} & 0.2 ($10^{-5}$) & 0.25 ($10^{-3}$) & 1 ($4\cdot10^{-7}$) & 1.5 ($10^{-5}$)\\
                \hline
                \textbf{CR-BCH(63,45)} & 0.2 ($10^{-5}$) & 0.25 ($10^{-4}$) & 0.75 ($2\cdot10^{-7}$) & 0.75 ($3\cdot10^{-6}$) \\
                \hline
                \textbf{CR-BCH(127,64)} & 0.3 ($10^{-4}$) & 0.4 ($10^{-3}$) & 0.75 ($10^{-6}$) & 1.25 ($10^{-4}$) \\
                \hline
            \end{tabular}
            \begin{tablenotes}
              \small
              \item (*) all gains are WRT the original WBP [1],[2].
             \item (**) The measured error value, where the gain is observed, is specified in parentheses.
            \end{tablenotes}
      \end{threeparttable}
\end{table*}

We simulate the WBP over a validation set of 1dB to 10dB until at least 1000 errors are accumulated at each given point. In addition, we adopt the syndrome based early termination, as we saw that some correctly decoded codewords were misclassified again by the following layers. This can also benefit complexity since the average number of iterations is less than or equal to 5 when using this rule.

Results for the simulation are presented in Figure \ref{fig:performance}. One can see that both distance-based and reliability-based approaches outperform the original BP-FF model with hyperparameters as in \cite{nachmani2016learning,nachmani2018deep}. We separate the contribution of our methods to two different regions. At the waterfall region the improvement varies from 0.25dB to 0.4dB in FER and 0.2dB to 0.3dB in BER for the different codes. At the error-floor region, the gain is increased by 0.75dB to 1.5dB in FER and by 0.75 to 1dB in BER for all codes. The best decoding gains per code are summarized in Table III. The measured error value, where the gain is observed, is specified in parentheses. Comparing to \cite{lian2019learned} in the BER graphs, a gain of 0.25dB is achieved in the CR-BCH(63,36) code, while in CR-BCH(127,64) one can observe similar performance. Furthermore, the difference in gains between the reliability curve and the reliability \& $d_H$ filtering curve indicates that the two methods indeed train on different distributions of words.

The FER metric is observed to gain the most from all approaches, with the reliability \& $d_H$ filtering approach having the best performance. One conjecture is that all these methods are optimized to improve FER directly. For the Hamming distance approach, lowering the number of errors in a single codeword reflects the FER directly. The reliability parameters are taken as a mean over the received words, thus adding more information on each $\mathbf{y}$ rather than on each single bit, $y_i$. One can see that all methods achieve better performance while keeping the same decoding complexity as before in \cite{nachmani2016learning,nachmani2018deep}. This is achieved solely by smartly sampling the data to train the neural decoder.

\section{Conclusions} \label{conclusions}

In this paper we proposed two novel sampling methods, incorporating error decoding measures with methodologies from the vast machine-learning field. Increases in performance of up to 0.4dB at the waterfall region, and of up to 1.5dB at the error-floor region, compared to the original WBP, are possible with no decoding complexity penalty, only by smartly sampling the training set. Furthermore, note that an aggregated increase in gain of about 2dB in high SNR, compared to BP, is achieved. We provided general guidelines for choosing training data in communications, starting in data exploration, validating assumptions by experiments and finally developing active learning based algorithms. We highlighted that SNR does not reveal the whole story. By introducing other key parameters one can have more control over the training data. Our conjecture is that sampling close to the decision boundary is crucial. At last, we urge the readers to seek sampling schemes in their communication application.

As for the next step, one may aim to find new ways of incorporating important parameters in training and validation for improved results. Likewise, one may explore a reinforcement learning algorithm which finds the optimal parameters during training with no conjectures whatsoever. Another direction is applying the proposed methods into the mRRD decoder \cite{nachmani2018near,dimnik2009improved} for approaching maximum-likelihood performance with further complexity reduction. Lastly, further analysis of dropped training samples could enhance explainability and provide insights about the proposed methods.

\section*{Acknowledgment}
We would like to thank Eran Asa for the insightful discussions. We also thank the reviewers and the editor for their beneficial comments.

\end{document}